\documentclass{revtex4}
\newcommand{\be}{\begin{equation}}
\newcommand{\ee}{\end{equation}}
\begin{document}
\title{Entropy and Entanglement in Quantum Ground States}
\author{M.~B.~Hastings}
\affiliation{Center for Nonlinear Studies and Theoretical Division,
Los Alamos National Laboratory, Los Alamos, NM, 87545}
\begin{abstract}
We consider the relationship between correlations and entanglement
in gapped quantum systems, with application to matrix product
state representations.
We prove that there
exist gapped one-dimensional local Hamiltonians such that
the entropy is exponentially large in the correlation length,
and we present
strong evidence supporting a conjecture that there exist such
systems with arbitrarily large entropy.
However, we then show that, under an assumption
on the density of states
which is believed to be satisfied by many physical systems
such as the fractional quantum Hall effect, that an efficient
matrix product state representation of the ground state
exists in any dimension.  Finally, we comment on the
implications for numerical simulation.
\end{abstract}
\maketitle

Finding the ground state of a local quantum Hamiltonian is one of
the basic problems in physics.  Efficient numerical techniques, however,
only exist for certain special cases.  Quantum Monte Carlo, for
example, is most effective for systems which lack a sign problem.  Fortunately,
in one dimension, the extremely powerful Density Matrix Renormalization
Group (DMRG)\cite{dmrg} algorithm is available, at least for systems with
an excitation gap.  This algorithm is based ultimately on knowledge of
the structure of the ground state: it is not an arbitrary wavefunction
but rather has the special form of a a matrix-product state\cite{mps}.
Recently, a very promising algorithm capable of finding higher-dimensional
matrix product states to approximate the ground state of fairly arbitrary
Hamiltonians has been developed\cite{peps}, and appears to offer, at
least for certain systems, the possibility of studying systems that
cannot be attacked by any other numerical technique.

All this work raises the important question: are ground states of such
gapped, local quantum systems indeed close to matrix product states?
A basic consequence of such a description is an {\it area law}: the
reduced density matrix of the ground state wavefunction on some
subvolume of the entire system has an entropy that is bounded
by some constant times the surface area of that volume, while critical
systems in one dimension may exhibit logarithmic corrections\cite{areal}.
One advance\cite{proj,kitaev} was to show that for all such Hamiltonians, the
Hamiltonian could be written as a sum of local terms such that the
ground state wavefunction was close to an eigenvector of each term
separately, allowing one to use certain generalized matrix product
states.  Unfortunately, these states required a hierarchical construction,
joining blocks of the system together at successively longer length scales,
and hence do {\it not} provide a construction of the
desired local matrix product states.  Further
this work left open the question of the number of
such states required in each block, precisely the question of
whether an area law holds or not.

Let $X,Y$ be sets of sites in the system.
An area law would follow from the assumption that the reduced
ground state density matrix of the system $\rho_{X\cup Y}$
on the set of sites $X\cup Y$
approximately factorizes into a product of density matrices,
$\rho_X \otimes \rho_Y$, if $X,Y$ are separated by some distance from each
other, as discussed in the Appendix.
As a partial result to showing this factorization,
it was shown that a gapped system with a local Hamiltonian
has exponentially decaying correlations\cite{lsm,loc}, and so
for any operators $O_X,O_Y$ with support on $X,Y$ we have
${\rm Tr}(O_X O_Y \rho_{X\cup Y}) \approx {\rm Tr}(O_X O_Y \rho_X \otimes
\rho_Y)$, up to exponentially small corrections in the distance
between $X$ and $Y$.  However,
the exponential decay of correlations functions is not directly
useful in showing the factorization of density matrices, as there
are so-called ``data hiding" states on bipartite systems for which
correlations are very small despite a high degree of
entanglement\cite{hiding,frank}.

In this paper, we study this question of the entanglement entropy,
for a gapped system with a local Hamiltonian.  We prove that
there exist one-dimensional systems for which this entropy is exponentially
large in the correlation length.  
Further, we present evidence to support a conjecture that for
any, arbitrarily large, $S$ there exist
one-dimensional systems with an entanglement entropy equal to $S$,
and with a Hilbert space dimension on each
site $D$ equal to $3$ or $4$ and a correlation length
bounded above by some $S$-independent constant of order unity.
Still, all these systems
involve long-range interactions so that the entropy is still
of order $\log(D)$ times the Lieb-Robinson group velocity\cite{lr}
divided by the energy gap.  

However, under a certain assumption on the density of
states believed to be satisfied by many important physical
systems such as the $1/3$ fractional Hall effect (in this
case defined for a lattice system of electrons), we
show that systems in arbitrary dimensions can be
efficiently represented as higher dimensional matrix
product states.  This result is based on a
recent result\cite{proj}, that the density matrix
at a nonzero temperature can be represented as a matrix product
density matrix\cite{mpop}.

\section{One Dimensional Model System and The Expander Graph State}
In this section, we construct the one dimensional model systems described
above.
Following\cite{mps,mps2}, we construct the system by first
writing its ground state as a matrix product, and then
defining the Hamiltonian as a sum of projection operators.
We write the ground state for this $V$ site system as
\be
\label{mps1}
\Psi(s_1,s_2,...,s_V)={\rm Tr}\Bigl(A(s_1) A(s_2)... A(s_V)\Bigr),
\ee
where $1\leq s_i\leq D$ is the state of the system on
site $i$ and where the $A(s_i)$ are $k$-by-$k$
dimensional matrices, with $k$ denoting the
{\it dimension} of the matrix product state.
We normalize so that
\be
\label{nmlze}
\sum_{s=1}^D A(s)A^{\dagger}(s)=1.
\ee
This state (\ref{mps1}) is the ground
state of a Hamiltonian,
${\cal H}=\sum_i P_{i,i+1,...,i+l}$ where
$P_{i,i+1,...,i+l}$ projects onto the
set of states $\Psi_{\alpha,\beta}$
on sites $i,i+1,...,i+l$
defined by $\Psi_{\alpha,\beta}(s_i,s_{i+1},...,s_l)=
\sum_{\{\gamma\}} A(s_i)_{\alpha\gamma_1}A(s_{i+1})_{\gamma_1\gamma_2}...
A(s_{i+l})_{\gamma_l\beta}$,
where $A(s_i)_{\alpha\gamma}$, with $1\leq \alpha,\gamma\leq k$,
are matrix elements of $A(s_i)$.  Here, the interaction length
$l$ is of order $\log_D(k)$\cite{mps,mps2}, and hence the Lieb-Robinson\cite{lr}
group velocity $v$ of this system is of order $l$.

A sufficient condition\cite{mps,mps2} for this Hamiltonian 
to
have a unique ground state with a gap is that
the linear map from $k$-by-$k$ matrices to
$k$-by-$k$ matrices
\be
{\cal E}(M)=\sum_{s=1}^D A(s)MA^{\dagger}(s)
\ee
have one nondegenerate eigenvalue equal to
unity, and then have a gap to the next
largest (in absolute value) eigenvalue.  From Eq.~(\ref{nmlze}),
the eigenvector with unit eigenvalue 
is proportional to the unit matrix.

We now propose a specific choice of the
$A(s)$ such that ${\cal E}(M)$ has a
gap in its spectrum and such that
$k$ is exponentially large in $D$.  We pick
\be
A(s)=\frac{1}{\sqrt{D}} U(s),
\ee
where $U(s)$ is a unitary matrix
depending on $s$.  Thus, Eq.~(\ref{nmlze})
is automatically satisfied.
We then pick $U(s)$ following two
different rules, one rule for
$1\leq s \leq D/2$ and one for
$D/2<s\leq D$.

For $1\leq s\leq D/2$, we pick $U(s)$
to be a diagonal matrix with matrix elements
\be
U(s)_{\alpha\beta}=
\delta_{\alpha\beta} F(\alpha,s)
\ee
where $F(\alpha,s)$
is some function such that $F(\alpha,s)=\pm 1$
for all $\alpha,s$.  We pick $F(\alpha,s)$ such
that for any two $\alpha,\beta$ with $\alpha\neq \beta$,
\be
\label{decorr}
|\sum_{s=1}^{D/2} F(\alpha,s) F(\beta,s)| \leq D/4.
\ee

The question arises: for which values of $k$
and $D$ is it possible to satisfy Eq.~(\ref{decorr})?
Define the vector $v_{\alpha}$ to be a
vector in a $D/2$ dimensional vector space by
$v_{\alpha}=(F(\alpha,1)/\sqrt{D/2},F(\alpha,2)/\sqrt{D/2},...,F(\alpha,s)/\sqrt{D/2})$.
Then, to satisfy Eq.~(\ref{decorr})
we must find $k$ vectors $v_{\alpha}$ in a $D/2$
dimensional vectors space such that
the inner product between $v_{\alpha}$ and $v_{\beta}$ is less
than $1/2$
for $\alpha\neq \beta$, and hence the angle between the
vectors $v_{\alpha},v_{\beta}$ is greater than $\phi=\cos^{-1}(1/2)$.
It is known\cite{conwaybook,shannon} that for large $D$, it is
possible to do this for
$k\leq \exp(c D/2)$ such vectors for some constant $c>0$, and
hence $k$ may be exponentially large in $D$.

For $D/2<s\leq D$, we pick $U(s)$
to be the matrix with matrix elements
\be
U(s)_{\alpha\beta}=
\frac{1}{k} \sum_{a=0}^{k-1}
\exp[2\pi i a (\alpha-\beta)/k]
\hat F(a,s),
\ee
and we pick the $\hat F$ such that
for any two $a,b$ with $a\neq b$,
\be
\label{decorrF}
|\sum_{s=D/2+1}^D \hat F(a,s) \hat F(b,s)| \leq D/4.
\ee
Again, this is possible to do so long as $k\leq \exp(cD/2)$.

Eq.~(\ref{decorr}) implies that off-diagonal
elements of $M$ are reduced by the map ${\cal E}(M)$ while
Eq.~(\ref{decorr}) implies that off-diagonal elements of
$M$ in the Fourier basis are also reduced by this map.
We now use this idea to show that the map ${\cal E}(M)$ does indeed have
a gap between the unit eigenvalue and the next largest eigenvalue.
The completely positive map ${\cal E}(M)$
is a Hermitian linear operator from
$k$-by-$k$ matrices to $k$-by-$k$ matrices and hence the
eigenvalue which is second largest in absolute value
can be found by taking the maximum over all traceless matrices $M$ with
${\rm Tr}(M^2)=1$
of $|{\rm Tr}(M{\cal E}(M))|$.  Let ${\cal E}_1(M)=\sum_{s=1}^{D/2}
A(s) M A^{\dagger}(s)$ and let ${\cal E}_2(M)=\sum_{s=D/2+1}^D A(s) M
A^{\dagger}(s)$ so that ${\cal E}(M)={\cal E}_1(M)+{\cal E}_2(M)$.
Let $M_d$ denote the diagonal
part of such a matrix $M$ and let $P_d={\rm Tr}(M_d^2)$.  Then,
from Eq.~(\ref{decorr}), we have $|{\rm Tr}(M{\cal E}_1(M))|\leq
P_d/2+(1-P_d)/4$.  Let $M_{f}$ be the matrix
with elements $M_{f}^{ab}=(1/k) 
\sum_{\alpha\beta}
M_{\alpha\beta}
\exp[-2\pi i(a \alpha-b \beta)/k]$ and
let $M_{df}$ denote the diagonal components of this matrix $M_f$ with
$P_{df}={\rm Tr}(M_{df}^2)$.
Then, from Eq.~(\ref{decorrF}) we have 
$|{\rm Tr}(M{\cal E}_2(M))|\leq
P_{df}/2+(1-P_{df})/4$.
However, for a traceless matrix $M$ with ${\rm Tr}(M^2)=1$ we have
$P_{df}\leq 1-P_d$.  Therefore, for any traceless matrix $M$ we have
$|{\rm Tr}(M{\cal E}(M))|\leq 3/4$, showing the existence of a gap as claimed.

In the case of this construction, we showed that the correlation length
is bounded by a constant, independent of $D$ and $k$.  For a sufficiently
long chain, the entanglement entropy between two halves of the chain
is equal to $\log_e(k)$.
One can represent each site on this system with $D$ states by $\log_2(D)$
sites with $2$ states on each site.  In this case, the correlation length
of the system is of order $\log_2(D)$, while the entropy is of order
$\log_e(\exp(cD/2))=cD/2$, which is exponentially large in the correlation
length as claimed.

This construction allows us to take $k$ exponentially large
in $D$ and still have a gap in the spectrum of ${\cal E}(M)$, as claimed.
We now present an alternative construction which gives an arbitrarily
large $k$, along with evidence that it also leads to a gap in
${\cal E}(M)$.
We take $D=4$ and for $s=1,2$ we take the matrices $U(s)$ to be of the form 
\be
U(s)=P(s) \sigma(s),
\ee
where
$\sigma(s)$ is a diagonal matrix which entries equal to $\pm 1$, and
$P(s)$ is a permutation matrix (that is, it has exactly one zero in
each row and column).  We then take $U(3)=U(1)^{\dagger}$ and
$U(4)=U(2)^{\dagger}$.  We set
$P(3)=P(1)^{\dagger}$ and $P(4)=P(2)^{\dagger}$.
We now fix $k$ at an arbitrary value and then
argue that by an appropriate choice of $P$ and $\sigma$ it is
possible to have a gap in ${\cal E}(M)$.

Let $P(s,\alpha)$ denote the value of $j$ such
that the matrix element
$P(s)^{\alpha\beta}=1$.  That is, $P(s,\alpha)$
is the result of applying the given
permutation to $\alpha$.  
The operator ${\cal E}(M)$ is block diagonal: acting
on a diagonal matrix it produces a diagonal matrix and acting on
an off-diagonal matrix it produces an off-diagonal matrix.  We study
the spectrum of ${\cal E}(M)$ in each block separately, starting with
the action on a diagonal matrix.  Let $M(\alpha,\beta)$ be the matrix
with matrix elements $M(\alpha,\beta)_{\sigma\tau}=\delta_{\alpha\sigma}
\delta_{\beta\tau}$.  We have
\be
\label{rw}
{\cal E}(M(\alpha,\alpha))=\frac{1}{4} \sum_{s=1}^4 M(P(s,\alpha),P(s,\alpha)).
\ee
Eq.~(\ref{rw}) can by represented by a diffusion process on a graph.
The graph has vertices labeled by $\alpha=1...k$, and with an undirected edge
from vertex $\alpha$ to $\beta$ if for some $s=1,2$ we have $P(s,\alpha)=\beta$ or $P(s,\beta)=\alpha$.
This graph has fixed coordination number equal to $4$,
although
some vertices may have more than one edge connecting them if
for some $s\neq t$ we have $P(s,\alpha)=P(t,\alpha)$\cite{double}.
Then, Eq.~(\ref{rw}) implies that in the first block, ${\cal E}(M)$ has
the same spectrum as the $1/4$ times the
adjacency matrix of the given graph.  However,
it is known that it is possible to find graphs with $k$ vertices
and fixed coordination
number $q$, for any $q>2$, such that $1/q$ times
the adjacency matrix has a unit
eigenvalue and then a gap to the next eigenvalue which is bounded below
by some $k$-independent constant\cite{graphlap}.
Indeed, it has been shown that the
gap between the unit eigenvalue and the next eigenvalue of the
adjacency matrix of a {\it random} graph generated using the
above procedure with a random choice of permutations $P$ is, with
probability tending to unity as the size $k$ of the graph tends to infinity,
bounded below by some $k$-independent constant\cite{graphlap2}.
Graphs with such a gap in the adjacency matrix spectrum
are called expander graphs and we refer to the state with the appropriate
choice of permutation matrices and diagonal matrices $\sigma(s)$ discussed
below as the {\it expander graph state}.

Now we consider the spectrum of ${\cal E}(M)$ in the second block.  We have
\be
\label{rw2}
{\cal E}(M(\alpha,\beta))=\frac{1}{4} \Bigl(
\sum_{s=1}^2 M(P(s,\alpha),P(s,\beta))\sigma(s)_{\alpha\alpha}\sigma(s)_{\beta\beta}+
\sum_{s=3}^4 M(P(s,\alpha),P(s,\beta))\sigma(s)_{P(s,\alpha)P(s,\alpha)}\sigma(s)_{P(s,\beta)P(s,\beta)}
\Bigr).
\ee
While it is possible to prove the existence of a gap in the spectrum
of ${\cal E}(M)$ in the first block for suitable matrices $P$,
we have to apply some physical intuition
to show that the second block has all eigenvalue separated from unity by
a constant.
If all the $\sigma(s)^{\alpha\alpha}$ were
equal to plus one, then the second block would
have one unit eigenvalue, with eigenvector proportional to a 
matrix with all off-diagonal
entries equal to unity and all diagonal entries equal to zero.  
In this case,
Eq.~(\ref{rw2}) would describe two correlated random walks on the given graph,
one for each index $\alpha,\beta$.
Then, for
a random choice of the permutations $P$, we would expect that, for similar
reasons to the existence of a gap in the diagonal sector, there would typically
be a gap to the next eigenvalue in the off-diagonal sector as any
correlations between $\alpha,\beta$ would be short-lived under this
random process.  We 
instead randomly
set each $\sigma(s)^{\alpha\alpha}$ equal to $\pm 1$, independently
for each $\alpha$.
In this case, we expect that random choices of the $\sigma$ and $P$ will,
due to the random signs, cause all eigenvalues in the second block to
be separated from unity by at least some $k$-independent
gap with probability tending to unity
as $k$ tends to infinity.

We have tested this numerically, by generating random permutations and random
$\sigma$.  We picked permutations with the additional restriction that
each permutation had exactly one cycle of length $k$.  We performed tests
on systems with $k$ up to $50$
and found that there was always one unit eigenvalue and a gap to the rest of
the spectrum, with the most negative eigenvalue separated by
a gap from $-1$, and further that the spectrum away from the unit eigenvalue
exhibited a scaling collapse, in that the density of eigenvalues appeared
to be roughly equal to $k^2$ times some $k$-independent function.

In the above, we considered $D=4$ so that we could take the matrix
$U(3)=U(1)^{\dagger}$ and $U(4)=U(2)^{\dagger}$ and arrive at a
real spectrum for ${\cal E}(M)$.  If we take $D=3$, and choose
$U(s)=P(s) \sigma(s)$ for $s=1,2,3$ with
$P(1),P(2),P(3)$ to be random permutations, and $\sigma(1),\sigma(2),\sigma(3)$
random diagonal matrices with entries $\pm 1$,
then for the first block of ${\cal E}(M)$ we arrive at the adjacency matrix
of a
{\it directed} graph of degree $3$.  In this case, ${\cal E}(M)$ may
have a complex spectrum.  However, we still expect there to be a gap.

\section{Matrix Product States from Thermal Density Matrices}

In this section, we build on the result in \cite{proj} that
it is possible, for local Hamiltonians to approximate the thermal density
matrix by a matrix product operator to show,
subject to an assumption on the density of states, that the
ground state is close to a matrix product state.
We start by considering the case of a unique ground state and a gap
$\Delta E$ to the next excited state and then generalize to multiple
ground states.
The idea is as follows: we approximate the thermal density matrix,
$Z^{-1} \exp[-\beta H]$ for a Hamiltonian $H$, with
$Z={\rm Tr}(\exp[-\beta H])$, by a matrix product density operator
up to some small error {\it in trace norm}.
For large enough $\beta$, the thermal density matrix becomes a good
approximation {\it in operator norm} to 
$\rho_0=\Psi_0\rangle\langle\Psi_0$, the projector
onto the ground state $\Psi_0$ of the given Hamiltonian.
We then make an assumption on the number of low energy states of the
Hamiltonian which then allows to show that the thermal density matrix
is a good approximation in trace norm to the projector, and hence that
our matrix product operator is a good approximation in trace norm to
the projector.  This means that in any complete orthonormal basis
there must exist some state such that the matrix product operator
acting on the basis is close to the ground state; picking this
basis to be a factorized basis gives us a matrix product state which
is close to the ground state.

We consider
a regular lattice (this condition can be weakened to include other
lattices) of $V$ sites labeled $1,2,..,V$, with a metric ${\rm d}(i,j)$ between
sites $i,j$ and we consider a Hamiltonian $H$ which is a sum of
terms $H_i$ supported on the sites within distance $R$ of $i$ for
some range $R$ and with operator norm $\Vert H_i \Vert$ bounded by some
constant $J$.  Then, it
was shown that for any $\epsilon$
and for any regular $d$-dimensional lattice of $V$ sites, there exists a matrix
product density operator approximation to the thermal
density matrix of the form
\be
\rho(\beta,l_{proj})=\sum_{\{\alpha_k\}} \rho_1(\alpha_1)
\rho_2(\alpha_2)...\rho_V(\alpha_V) F_1(\{\alpha_j\}) F_2(\{\alpha_j\})
... F_V(\{\alpha_j\}),
\ee
where ${\rm Tr}(\rho(\beta,l_{proj}))=1$,
where each operator $\rho_i(\alpha_i)$ acts only on site $i$,
where the $\alpha_j$ are some set of indices, with range $1\leq \alpha_j \leq \alpha_{max}$ for all $j$,
and where
each function $F_i$ depends only on the $\alpha_j$ with ${\rm d}(i,j)\leq l_{proj}$,
and such that
\be
{\rm Tr}(|\rho(\beta,l_{proj})-Z^{-1} \exp[-\beta H]|)
\leq \epsilon
\ee
where ${\rm Tr}(|...|)$ denotes the trace norm, and where
\be
l_{proj} \sim R \log(V\beta/J\epsilon),
\ee
with the constant of proportionality depending on the exact lattice
structure,
and with
\be
\alpha_{max} \sim D^{l_{proj}^d \beta/J}.
\ee

Now, we want to introduce a condition on the density of states of
a given Hamiltonian that will lead to a bound on how accurately the
thermal density matrix approximates the ground state projection
operator.  To motivate this bound,
suppose that we have a Hamiltonian of the following particularly
simple form: $H=\Delta E \sum_i O_i$, where $O_i$ is an operator
on site $i$ with one zero eigenvalue and $D-1$ eigenvalues equal to unity.
Then, there exists one state, with all spins down, with energy $0$.
There exist $(D-1) V$ states, each with one spin up, with energy $\Delta E$.
There exist $(D-1)^2 V(V-1)/2$ states with energy $\Delta E/2$, and so on.
Thus, we have the bound that
$\rho(m)$, which we define to be
the number of states with energy $E_i$ with $m\Delta E\leq E_i<
(m+1) \Delta E$, obeys
\be
\label{assump}
\rho(m)\leq (c V)^m/m!,
\ee
for $c=D-1$.

In general, however, many gapped systems will obey assumption (\ref{assump})
for some constant $c$.  All free fermion systems with a gap in the
single particle spectrum will obey this assumption.  Physically,
we expect that a lattice realization of the fractional Hall effect at
an incompressible filling fraction will also obey this.  In the rest
of this section, we show that if a system does obey this assumption,
then it is possible to write the ground state as a higher-dimensional
matrix product state,
providing bounds on the error as a function of the dimension of the matrix
product state.
Using this assumption,
\be
{\rm Tr}(|Z^{-1}\exp[-\beta H]-\rho_0|)\leq
2 \sum_{m=1}^{\infty} (\exp[-\beta \Delta E] c V)^m/m!
=
2(\exp\{\exp[-\beta \Delta E]c V\}-1),
\ee
and hence for any $\epsilon$, for
$\beta\geq \log(c V/\log(1+\epsilon/2))/\Delta E$, we have
\be
{\rm Tr}(|Z^{-1}\exp[-\beta H]-\rho_0|)\leq \epsilon.
\ee

The operator $\rho(\beta,l_{proj})$ can be written in the form
$\rho(\beta,l_{proj})=O^{\dagger}O$ for some operator $O$.
We write an orthonormal basis of states for the system by vectors
$\Psi_{\vec v}\rangle$, where $\vec v=(v_1,v_2,...,v_V)$ with
$1\leq v_i \leq D$.
The vector $\Psi_{\vec v}\rangle$
denotes the state where site $i$ is in state $v_i$.  This is a factorized
basis of states for the system.
Then, $\rho_0=\sum_{\{\vec v\}} O^{\dagger} \Psi_{\vec v}\rangle\langle
\Psi_{\vec v} O$.
Let 
\be
p^0_{\vec v}=
\langle \Psi_{\vec v}, O \rho_0 O^{\dagger} \Psi_{\vec v} \rangle.
\ee
and
\be
p^>_{\vec v}=
\langle \Psi_{\vec v}, O O^{\dagger} \Psi_{\vec v} \rangle-p^0_{\vec v}.
\ee
Thus, $p^0_{\vec v}$ is equal to $|O^{\dagger}\Psi_{\vec v}|^2$ times the
probability that the state $O^{\dagger}\Psi_{\vec v}/|O^{\dagger} \Psi_{\vec v}|$ 
is in the ground state $\Psi_0$,
with $p^>_{\vec v}$ is equal to $|O^{\dagger}\Psi_{\vec v}|^2$ times the
probability that the state is not in the ground state.
Then $\sum_{\vec v}(p^0_{\vec v}+p^>_{\vec v})=1$ and
$\epsilon\geq {\rm Tr}(|\rho-\rho_0|)\geq (1-\sum_{\vec v} p^0_{\vec v})+
\sum_{\vec v}p^>_{\vec v}$,
so $\sum_{\vec v}p^>_{\vec v}\leq \epsilon/2$.  Hence,
\be
\frac{\sum_{\vec v} p^>_{\vec v}}{\sum_{\vec v} p^0_{\vec v}}
\leq
\frac{\epsilon/2}{1-\epsilon/2}.
\ee
Hence, there must exist some $\vec w$ such that
\be
\frac{p^>_{\vec w}}{p^0_{\vec w}}
\leq
\frac{\epsilon/2}{1-\epsilon/2}.
\ee

Then,
\be
\Psi_{mps}\equiv \frac{1}{\sqrt{p^0_{\vec w}+p^>_{\vec w}}}
O^{\dagger}\Psi_{\vec w}\rangle
\ee
is a normalized matrix product state,
by the assumption that $O$ is a matrix product operator.
Further,
\be
\label{error}
|\langle \Psi_0,\Psi_{mps} \rangle|^2=
\frac{p^0_{\vec w}}{p^0_{\vec w}+p^>_{\vec w}}
\geq \frac{1}{1+(\epsilon/2)/(1-\epsilon/2))}=1-\epsilon/2+{\cal O}(\epsilon^2),
\ee
so that $\Psi_{mps}$ is close to the state $\Psi_0$.

Thus, for any $\epsilon$, we can find a matrix product state
$\Psi_{mps}$ such that Eq.~(\ref{error}) holds and such that 
$l_{proj}$ grows logarithmically in $V$ with $\alpha_{max}$ growing
as an exponential of a polylog in $V$.

This derivation can be readily generalized to the case of multiple
ground states, labeled $\Psi_0^{(a)}$, $a=0...n-1$.  Let $\rho_0=
n^{-1}\sum_{a=1}^n \Psi_0^a\rangle\langle\Psi_0^a$, and assume that
there exists a $\rho=O^{\dagger}O$ with ${\rm Tr}(\rho)=1$ and
${\rm Tr}(|\rho-\rho_0|)\leq \epsilon$.  Then, define $p_{\vec v}^0$
as above and again find an appropriate $\Psi_{\vec w_0}$, so that
$\Psi_{mps,0}=(1/\sqrt{p_{\vec w_0}^0+p_{\vec w_0}^>}) O^{\dagger} 
\Psi_{\vec w_0}$
is close to some state $\Psi_0^{(0)}$ in the ground state subspace.
Then, define $\rho_1=
\rho_0-n^{-1} \Psi_0^{(0)}\rangle\langle \Psi_0^{(0)}$, and
define
$p_{\vec v}^1=
\langle \Psi_{\vec v}, O \rho_1 O^{\dagger} \Psi_{\vec v} \rangle$.
One can then find a state $\Psi_{\vec w_1}$ such that
$\Psi_{mps,1}=(1/\sqrt{p_{\vec w_1}^1+p_{\vec w_1}^>}) O^{\dagger}
\Psi_{\vec w_1}$ is close to some state $\Psi_0^{(1)}$ in the ground
state subspace, with $\langle \Psi_0^{(1)},\Psi_0^{(0)} \rangle=0$.
Repeating this $n$ times, one can then span the ground state subspace with
a set of
matrix product states, $\Psi_{mps,a}$, $a=0...n-1$.

We note that in the procedure discussed in this section,
the thermal density matrix is introduced simply
as a means of approximating the projection operator by a local
operator.  Other approximate projection operators could have been used,
such as
\be
\frac{1}{\sqrt{2\pi}t_q}
\int_{-\infty}^{\infty} {\rm d}t \exp[i H t]\exp[-(t/t_q)^2/2],
\ee
for some $t_q$,
where the approximation becomes more accuracte, but also less local,
as $t_q$ increases.  In this case, given the entropy
assumption (\ref{assump}) we would need to take $t_q$ of
order $\log(V)/\Delta E$ to ensure that the approximate projection operator
was close in trace norm to the exact projection operator.

\section{Discussion}

We have shown that it is possible to find systems for which the
entanglement entropy between a given subvolume and the rest of the
system is much larger than expected from the correlation length.
However, we have also shown that, subject to the assumption (\ref{assump})
which is satisfied by many physical systems, matrix products give
a good representation of the ground state even in higher dimensional
systems, providing additional reason to consider the use of these
states as a numerical technique.

Eq.~(\ref{assump}) has an interesting physical interpretation.  If there is
a gap and this
equation is satisfied, then the system is not ``glassy" in that
at temperatures of order $\Delta E/\log(V)$ the system is in its
ground state with probability of order unity.
Consider, however, the following
system, related to an example of Terhal and DiVincenzo: a one-dimensional
system with $D=3$, represented by a spin-1 degree of freedom on each of
$V$ different sites.
The Hamiltonian is $-\sum_i [(S^z_i)^2-1/2] [(S^z_{i+1})^2-1/2]
+(1/V) \sum_i (S^z_i)^2$.  The ground state has spins with $S^z=0$ and
energy equal to $-V/4$.  However,
there are $2^V$ states with each spin having $S^z=\pm 1$ with energy $-V/4+1$.
Thus, this system does not satisfy Eq.~(\ref{assump}) for any fixed, 
$V$-independent, $c$.  However, the related Hamiltonian
$-\sum_i [(S^z_i)^2-1/2] [(S^z_{i+1})^2-1/2]
+\sum_i (S^z_i)^2$ has the same ground state and does satisfy 
Eq.~(\ref{assump}).  In general, for any classical system, meaning that
the Hamiltonian is a some of operators which are diagonal in some factorized
basis, the ground state will trivially be a matrix product state.
Thus, we conjecture that for any gapped local Hamiltonian it is possible
to find a related Hamiltonian with close to the same ground state and
which satisfies Eq.~(\ref{assump}) and hence has an approximate
matrix product ground state.

{\it Acknowledgments---}
I thank F. Verstraete for useful discussions.
This work was carried out under the auspices of the NNSA
of the U.S. DOE at LANL under Contract No. DE-AC52-06NA25396.

\appendix
\section{Density Matrix Factorization and an Area Law}

The goal in this section is to relate the factorization of the
density matrix to the entropy of the reduced density matrix.
Suppose $X$ is an arbitrary set and let $Y$ be the set of all sites
$i$ such that ${\rm dist}(X,i)\geq l$ for some $l$.
Let $B$ denote the set of sites $j$ which are neither in
$X$ nor in $Y$.
In this section, we assume that for some state $\Psi_0$ there is
a bound 
\be
\label{assumedbd}
{\rm Tr}(|\rho_{XY}-\rho_X\otimes\rho_Y|)\leq \epsilon
\ee
for some $\epsilon$ for the given
$X,Y$ and
derive a bound on the entropy of the reduced density matrix $\rho_X$.
The case that $\epsilon=0$ was considered by \cite{frank}, where it
was shown that this implies that the density matrix $\rho_X$ has
at most $D^{|B|}$ nonzero eigenvalues.  In this appendix, we
consider the case of a non-zero $\epsilon$.

The wavefunction $\Psi_0$ can be written as
\be
\Psi_0\rangle
=\sum_{\alpha} A(\alpha) \
\Psi_{B}^{\alpha}\rangle\otimes\Psi_{XY}^{\alpha}\rangle,
\ee
where $\Psi_{XY}$ is a wavefunction on $X \cup Y$ and $\Psi_{B}$ is
a wavefunction on the set $B$, where $B$ is
the set of all sites $i$ such that $i \not \in X$ and $i \not \in Y$
and where $\langle \Psi_{B}^{\alpha},\Psi_{B}^{\beta} \rangle=
\langle \Psi_{XY}^{\alpha},\Psi_{XY}^{\beta} \rangle=\delta_{\alpha,\beta}$.
Then,
\be
\rho_{XY}=\sum_{\alpha} |A(\alpha)|^2
\Psi_{XY}^{\alpha}\rangle\langle\Psi_{XY}.
\ee

Then, $\rho_{XY}$ is equal to the weighted
sum of at most $D^{|B|}$ different
density matrices
$\Psi_{XY}^{\alpha}\rangle\langle\Psi_{XY}$, each of which corresponds to
a pure state.
Define $P$ to be the projection operator onto the space spanned by
these $D^{|B|}$ different states $\Psi_{XY}^{\alpha}$.
Then, from Eq.~(\ref{assumedbd}), we have
${\rm Tr}((\rho_{XY}-\rho_X\otimes\rho_Y)P)\leq \epsilon$ so therefore
${\rm Tr}(P \rho_X\otimes\rho_Y)\geq 1-\epsilon$.

The operator $\rho_X$ may be diagonalized, so that
\be
\rho_X=\sum_{\alpha}\rho_X(\alpha) \Psi_X^\alpha\rangle\langle\Psi_X^\alpha,
\ee
where $\langle \Psi_X^{\alpha},\Psi_X^{\beta} \rangle=\delta_{\alpha,\beta}$.
The index $\alpha$ ranges from $1$ to at most $D^{|X|}$.  Let us
order the eigenvalues, so that
$\rho_X(\alpha)\geq \rho_X(\beta)$ if $\alpha>\beta$.
Similarly, we write $\rho_Y=\sum_{\alpha}=\rho_Y(\alpha)
\Psi_Y^\alpha\rangle\langle\Psi_Y^\alpha$ and again order the
eigenvalues so that
$\rho_Y(\alpha)\geq \rho_Y(\beta)$ if $\alpha>\beta$.
Finally we denote the eigenvalues of $\rho_X \otimes \rho_Y$ by
$\rho(\gamma)$ and also order these eigenvalues.
Note that for each $\gamma$, $\rho(\gamma)=\rho(\alpha)
\rho(\beta)$ for some $\alpha,\beta$ with $\alpha\leq \gamma$.
Since $P$ projects onto a space with dimension at most $D^{|B|}$,
we have 
\be
1-\epsilon\leq
{\rm tr}(P \rho_X \otimes \rho_Y)\leq \sum_{\gamma=1}^{D^{|B|}}
\rho(\gamma)
\leq \sum_{\alpha=1}^{D^{|B|}} \rho_X(\alpha).
\ee
Therefore,
\be
\label{few}
\sum_{\alpha=D^{|B}+1}^{\alpha=D^{|X|}} \rho_X(\alpha) \leq \epsilon.
\ee

Eq.~(\ref{few}) means that the total probability of the system
being in any state on set $X$
{\it other} than some given set of $D^{|B|}$ states is bounded by $\epsilon$.
This is very close to a result for the entropy.
Indeed, $\rho_X$ has at most $D^{|X|}$ non-zero eigenvalues.
Thus, the entropy is bounded by:
\be
S(\rho_X)\equiv -\sum_{\alpha} 
\rho_X(\alpha) \log(\rho_{\alpha})
=-\sum_{\alpha=1}^{D^{|B|}} 
\rho_X(\alpha) \log(\rho_{\alpha})-
\sum_{\alpha={D^{|B|}+1}}^{D^{|X|}}
\rho_X(\alpha) \log(\rho_{\alpha})
\leq |B| \log(D)+\epsilon  |X| (\log(D)+\log(\epsilon)).
\ee


\begin{thebibliography}{99}
\bibitem{dmrg} S. R. White, Phys. Rev. Lett. {\bf 69}, 2863 (1992).

\bibitem{mps} M. Fannes, B. Nachtergaele, and R. F. Werner, Commun. Math. Phys.
{\bf 144}, 443 (1992).

\bibitem{peps} F. Verstraete and J. I. Cirac, preprint/cond-mat/0407066.

\bibitem{areal} G. Vidal, J. I. Latorre, E. Rico, and A. Kitaev, Phys. Rev.
Lett. {\bf 90}, 227902 (2003); F. Verstraete, M. M. Wolf, D. Perez-Garcia, and J. I.
Cirac, Phys. Rev. Lett. {\bf 96} 220601 (2006).

\bibitem{proj} M. B. Hastings, Phys. Rev. B {\bf 73}, 085115 (2006).

\bibitem{kitaev} A. Kitaev, preprint cond-mat/0506438.

\bibitem{lsm} M. B. Hastings,
Phys. Rev. B {\bf 69}, 104431 (2004).

\bibitem{loc} M. B. Hastings, Phys. Rev. Lett. {\bf 93}, 140402 (2004).

\bibitem{frank} F. Verstraete and J. I. Cirac, Phys. Rev. B {\bf 73},
094423 (2006).

\bibitem{hiding} P. Hayden, D. Leung, P. W. Shor, and A. Winter,
Commun. Math. Phys. {\bf 250(2)}, 371 (2004).

\bibitem{mpop} The operators we will consider are related to those
in F. Verstraete, J. J. Garc\'{i}a-Ripoll, and J. I. Cirac,
Phys. Rev. Lett. {\bf 93}, 207204 (2004), although we consider them
in arbitrary dimension and hence rather than a matrix product
they are represented by a sum over bond variables.

\bibitem{mps2} D. P\'{e}rez-Garc\'{i}a, F. Verstraete, M. M. Wolf, and
J. I. Cirac, quant-ph/0608197.

\bibitem{lr} E. H. Lieb and D. W. Robinson, Commun. Math. Phys. {\bf 28}, 
251 (1972);
M.  B. Hastings and T. Koma,
Commun. Math. Phys. {\bf 265}, 781 (2006);
B. Nachtergaele and R. Sims, Commun. Math. Phys.
{\bf 265}, 119 (2006).


\bibitem{conwaybook} J. H. Conway and N. J. A. Sloane, {\it Sphere
Packing, Lattices and Groups} (Springer-Verlag, New York 1988).

\bibitem{shannon} C. E. Shannon, Bell Syst. Tech. Jour. {\bf 38}, 611 (1959).

\bibitem{graphlap} N. Alon, Combinatorica, {\bf 6(2)}, 83 (1986).

\bibitem{double} With the random choice of permutations below, typically
only $O(1)$ nodes will have coordination number different from $4$.

\bibitem{graphlap2}
J. Friedman, Conf. Proc. of the Annual ACM
Symposium on Theory of Computing, 720 (2003); J. Friedman, preprint cs/0405020.
\end{thebibliography}
\end{document}